\documentclass[twocolumn]{aastex631}

\begin{document}

\title{Little Red Dots or Brown Dwarfs?\\NIRSpec Discovery of Three Distant Brown Dwarfs Masquerading as NIRCam-Selected\\Highly-Reddened AGNs}

\correspondingauthor{Danial Langeroodi}
\email{danial.langeroodi@nbi.ku.dk}

\author[0000-0001-5710-8395]{Danial Langeroodi}
\affil{DARK, Niels Bohr Institute, University of Copenhagen, Jagtvej 128, 2200 Copenhagen, Denmark}

\author[0000-0002-4571-2306]{Jens Hjorth}
\affil{DARK, Niels Bohr Institute, University of Copenhagen, Jagtvej 128, 2200 Copenhagen, Denmark}

\begin{abstract}

Cold, substellar objects such as brown dwarfs have long been recognized as contaminants in color-selected samples of active galactic nuclei (AGNs). In particular, their near- to mid-infrared colors (1--5 $\mu$m) can closely resemble the V-shaped ($f_{\lambda}$) spectra of highly-reddened accreting supermassive black holes (``little red dots''), especially at $6 < z < 7$. Recently, a NIRCam-selected sample of little red dots over 45 arcmin$^2$ has been followed up with deep NIRSpec multi-object prism spectroscopy through the UNCOVER program. By investigating the acquired spectra, we identify three of the 14 followed-up objects as T/Y dwarfs with temperatures between 650 and 1300 K and distances between 0.8 and 4.8 kpc. At $4.8^{+0.6}_{-0.1}$ kpc, Abell2744-BD1 is the most distant brown dwarf discovered to date. We identify the remaining 11 objects as extragalactic sources at $z_{\rm spec} \gtrsim 5$. Given that three of these sources are strongly-lensed images of the same AGN (Abell2744-QSO1), we derive a brown dwarf contamination fraction of 25\% in this NIRCam-selection of little red dots. We find that in the near-infrared filters, brown dwarfs appear much bluer than the highly-reddened AGN, providing an avenue for distinguishing the two and compiling cleaner samples of photometrically selected highly-reddened AGN.

\end{abstract}


\section{Introduction} \label{sec: intro}

JWST photometry has revealed a population of red and extremely compact sources, the little red dots, representing faint and/or highly-reddened active galactic nuclei (AGNs) at $z > 3$ \citep{2023MNRAS.524.2312E, 2023ApJ...952..142F, 2023ApJ...942L..17O, 2023MNRAS.525.1353J, 2023arXiv230515458B, 2023arXiv230514418B, 2023ApJ...950L...5Y, Labbe+2023}. The nature of some of these sources is being confirmed through NIRCam wide field slitless spectroscopy \citep{2023arXiv230605448M} and NIRSpec follow-up campaigns \citep{2023Natur.619..716C, 2023arXiv230311946H, 2023arXiv230200012K, 2023arXiv230308918L, 2023arXiv230206647U, 2023arXiv230514418B, 2023arXiv230512492M, 2023arXiv230801230M, 2023arXiv230805735F}. Constraining the number density of faint AGN at these high redshifts can be fundamental for distinguishing between different massive black hole formation and growth scenarios as well as understanding their role in reionizing the Universe. Moreover, constraining the properties of their host galaxies can be crucial in calibrating the massive black hole-host galaxy co-evolution models, and in particular, the role of AGN feedback in regulating star formation. 

Samples of tens of faint AGNs at $z > 3$ are being confirmed spectroscopically \citep{2023arXiv230311946H, 2023arXiv230605448M, 2023arXiv230801230M}, enabling tremendous progress in constraining their number density. However, due to small survey areas, insufficient depths, or the complex selection functions of spectroscopic surveys, tight constraints on the luminosity function of faint and/or highly reddened AGN might prove challenging without fully leveraging wide photometry-only surveys of NIRCam and MIRI. Inferring the luminosity function of broad-band color-selected AGN is subject to contaminants, especially in the form of Galactic, cold, substellar objects, such as brown dwarfs \citep[see, e.g.,][]{Kirkpatrick+2011, 2012ApJ...753...30S, 2014MNRAS.439.1038W}. Depending on the available filters, the 1$\mu$m spectral features of brown dwarfs can closely mimic a Ly$\alpha$ break at $z = 6$--7.5 \citep{2014MNRAS.439.1038W, 2015MNRAS.452.1817B, hainline+2023}. Moreover, their V-shaped spectra, driven by temperature-dependent broad features at $\sim 1\mu$m and $\sim 4\mu$m \citep[see, e.g.,][]{2014ASInC..11....7B}, can resemble the V-shaped spectra of highly-reddened AGN at $z \sim 7$, characterized by a blue continuum at observed 1--3 $\mu$m and a red continuum at 3--5 $\mu$m \citep[see, e.g.,][]{Labbe+2023, 2023ApJ...952..142F, 2023arXiv230805735F}.

Fully leveraging the wide survey areas, depths, and simple selection criteria of photometry-only surveys in constraining the cosmic distribution and redshift evolution of red AGNs requires an accurate understanding of the prevalence of potential contaminants. In this work, we analyze the NIRSpec multi-object prism follow-up \citep[UNCOVER program;][]{Bezanson+2022} of a NIRCam-selected sample of 13 highly-reddened AGN \citep{Labbe+2023} to investigate the fraction of brown dwarf contaminants. We confirm three objects as brown dwarfs, and the remaining ten as extragalactic sources. Given that three of the extragalactic spectra correspond to the triply imaged AGN Abell2744-QSO1 \citep{2023ApJ...952..142F, 2023arXiv230805735F}, we infer a brown dwarf contamination fraction of $25\%$. By comparing the distribution of these objects on near- and mid-infrared color-color diagrams, we find that brown dwarfs are on average much bluer in near-infrared filters than extragalactic sources. Hence, simple color cuts can potentially minimize the contamination of brown dwarfs in color-selected samples of faint/highly-reddened AGN.

So far, only a handful of brown dwarfs have been targeted/discovered with JWST. These include the serendipitous discovery of a late T-type at $\sim 0.6$--0.7 kpc in NIRCam photometry \citep{2023ApJ...942L..29N}; NIRCam photometry of the late T-type HD 19467 B at $\sim 0.03$ kpc \citep{2023ApJ...945..126G}; discovery of a Y+Y-type brown dwarf binary at $\sim 0.01$ kpc in NIRCam photometry \citep{2023ApJ...947L..30C}; discovery of a T-type at $\sim 2.5$ kpc in NIRCam photometry \citep{2023MNRAS.523.4534W}; and the NIRSpec and MIRI spectroscopy of a Y-type at $\sim 0.014$ kpc \citep{2023ApJ...951L..48B}, marking the first JWST spectrum of a nearby brown dwarf.

In this work, we report the first NIRSpec spectra of three distant brown dwarfs. One of these spectra (Abell2744-BD3) confirms the brown dwarf discovered in NIRCam photometry by \cite{2023ApJ...942L..29N}. By fitting their spectral energy distributions, we infer effective temperatures of 1300 K, 1100 K, and 650 K for the early T-type Abell2744-BD1, mid T-type Abell2744-BD2, and the late T-type Abell2744-BD3, respectively. We infer distances of $\sim 4.8$ kpc, $3.0$ kpc, and $0.8$ kpc for Abell2744-BD1, -BD2, and -BD3 respectively. Abell2744-BD1 at 4.8 kpc is the farthest spectroscopically confirmed brown dwarf to date.

\vspace{20pt}
\section{Data} \label{sec: data}

\begin{deluxetable}{llll}
\tablewidth{0pt}
\tablecaption{Measured photometry of the brown dwarfs discovered in this work \citep[all magnitudes are in the AB system;][]{1983ApJ...266..713O}}
\label{table: photometry}
\tablehead{
\colhead{filter} &
\colhead{{\scriptsize Abell2744-BD1}} &
\colhead{{\scriptsize Abell2744-BD2}} &
\colhead{{\scriptsize Abell2744-BD3}\tablenotemark{a}}
}
\startdata
MSA ID & 32265 & 33437 & 39243 \\
RA\tablenotemark{b} & 3.537529 & 3.546420 & 3.513891 \\ 
Dec\tablenotemark{b} & $-30.370169$ & $-30.366245$ & $-30.356024$ \\
F090W & $-$ & $-$ & $31.84 \pm 1.99$ \\
F115W & $28.08 \pm 0.10$ & $27.62 \pm 0.10$ & $28.08 \pm 0.10$ \\
F150W & $27.93 \pm 0.10$ & $28.33 \pm 0.10$ & $28.90 \pm 0.10$ \\
F200W & $27.97 \pm 0.10$ & $29.07 \pm 0.10$ & $29.73 \pm 0.14$ \\
F277W & $28.38 \pm 0.10$ & $29.36 \pm 0.11$ & $29.29 \pm 0.10$ \\
F356W & $27.51 \pm 0.10$ & $28.40 \pm 0.10$ & $27.85 \pm 0.10$ \\
F410M & $26.94 \pm 0.10$ & $26.98 \pm 0.10$ & $-$ \\
F444W & $27.24 \pm 0.10$ & $27.28 \pm 0.10$ & $25.79 \pm 0.10$ \\
\enddata
\tablenotetext{a}{This is the brown dwarf discovered in NIRCam photometry by \cite{2023ApJ...942L..29N}.}
\tablenotetext{b}{J2000.0 (deg)}
\end{deluxetable}

% \begin{figure*}
%     \centering
%     \includegraphics[width=17cm]{brown_dwarfs.pdf}
%     \caption{Three out of the 15 little red dots followed up with deep NIRSpec multi-object prism spectroscopy turn out to be brown dwarfs. This figure shows the extracted 1D spectra of these brown dwarfs with the MSA ID of each object noted on the top of its corresponding panel. Abell2744-BD3 (bottom panel) is the brown dwarf identified in GLASS NIRCam photometry by \cite{2023ApJ...942L..29N}.}
%     \label{fig: brown dwarfs}
% \end{figure*}

\begin{figure*}
    \centering
    \includegraphics[width=17cm]{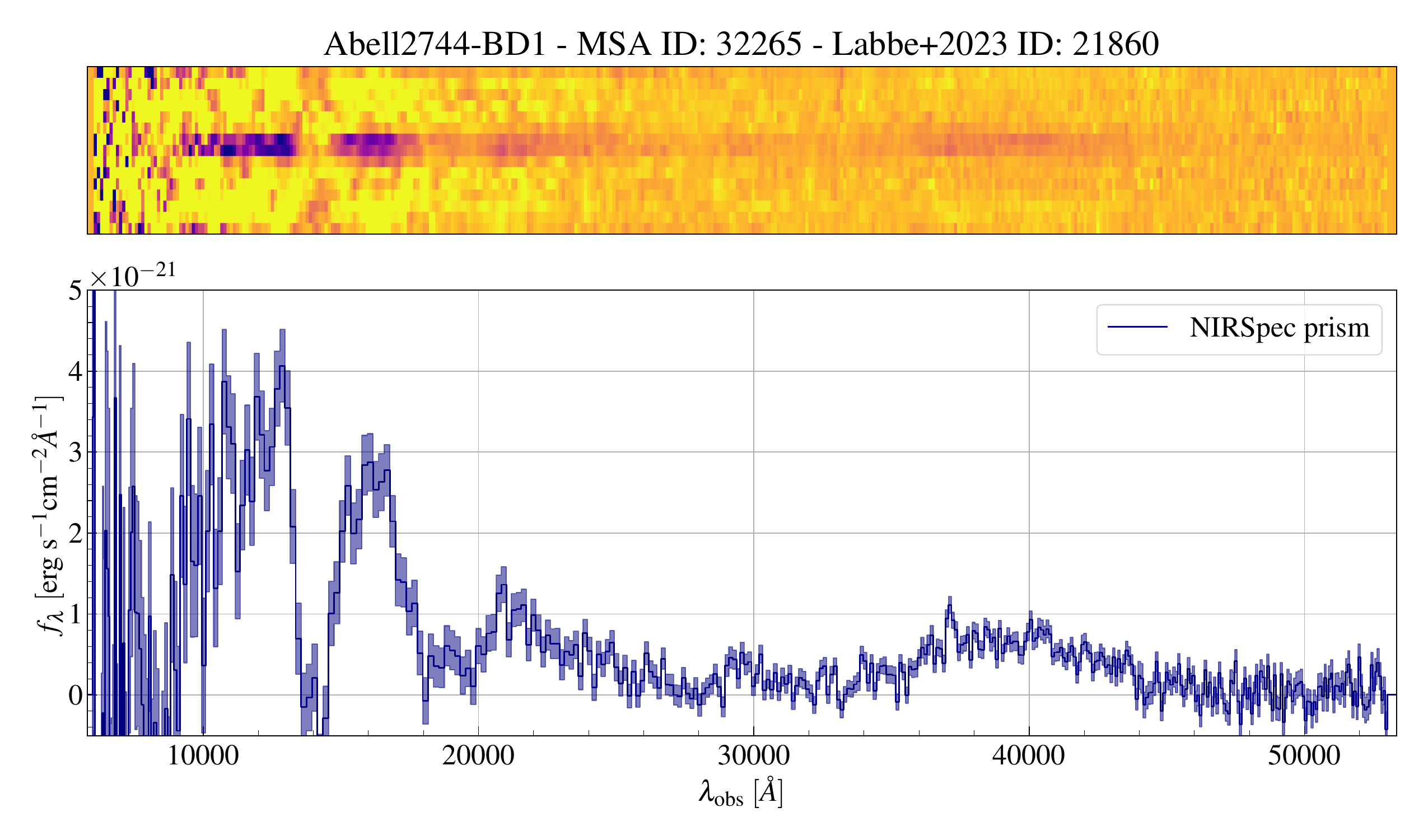}
    \caption{Three out of the 15 little red dots followed up with deep NIRSpec multi-object prism spectroscopy turn out to be brown dwarfs. This figure shows the 2D spectrum and the extracted 1D spectrum of the brown dwarf Abell2744-BD1.}
    \label{fig: spectrum BD1}
\end{figure*}

\begin{figure*}
    \centering
    \includegraphics[width=17cm]{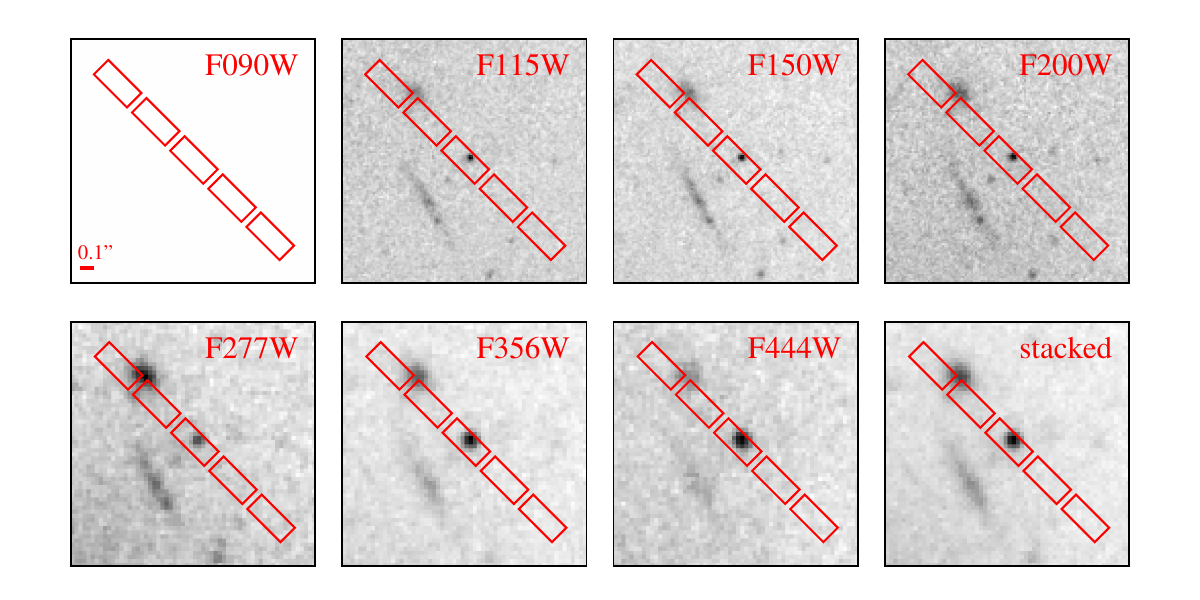}
    \caption{Photometry of Abell2744-BD1. Each inset presents a $2.4\arcsec\times2.4\arcsec$ cutout centered on the MSA slit that targeted the source. The bottom-right inset shows the variance-weighted stack of F277W, F356W, and F444W images. The MSA slitlets are overlaid as the red rectangles.}
    \label{fig: photometry BD1}
\end{figure*}

\begin{figure*}
    \centering
    \includegraphics[width=17cm]{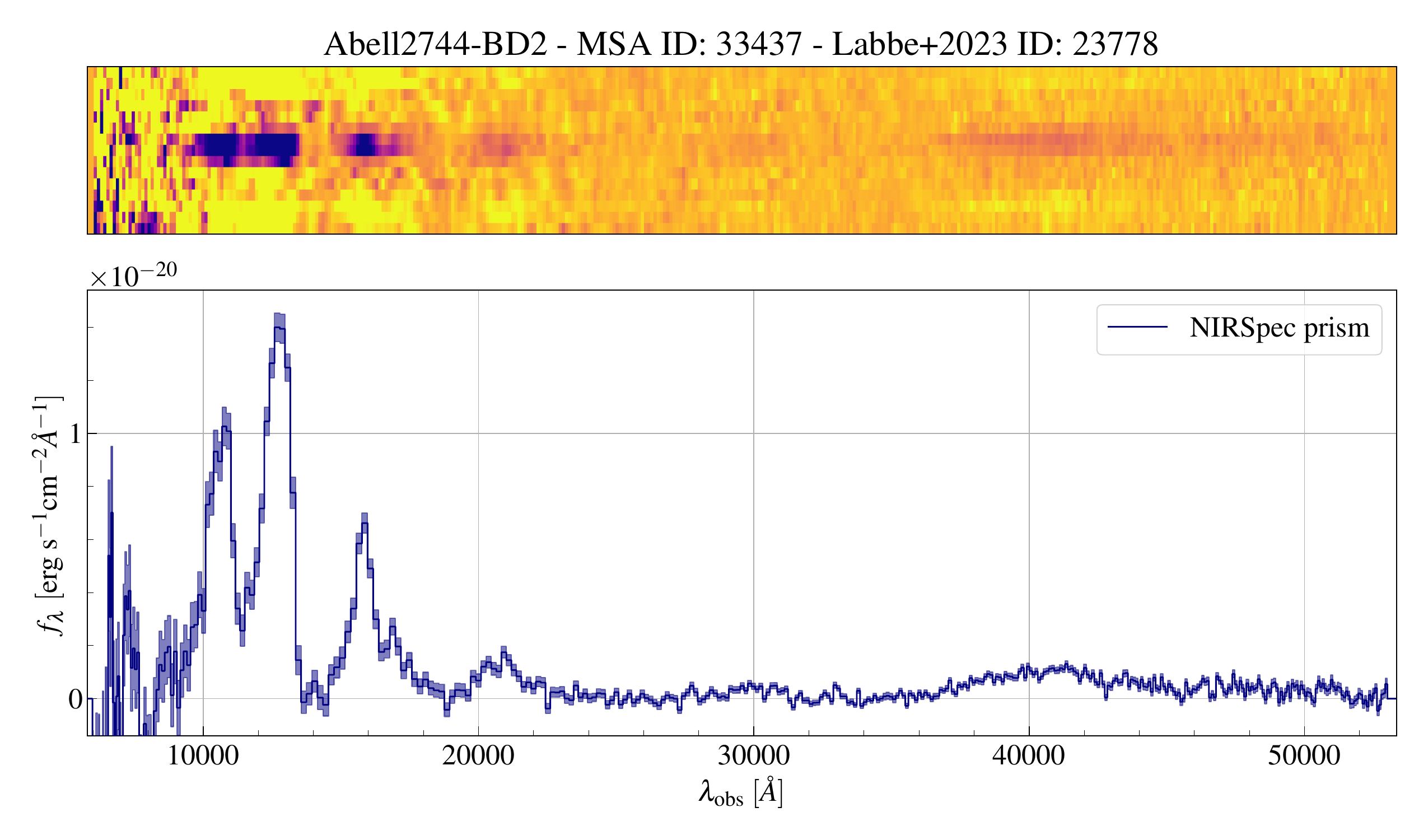}
    \caption{2D spectrum and the extracted 1D spectrum of the brown dwarf Abell2744-BD2.}
    \label{fig: spectrum BD2}
\end{figure*}

\begin{figure*}
    \centering
    \includegraphics[width=17cm]{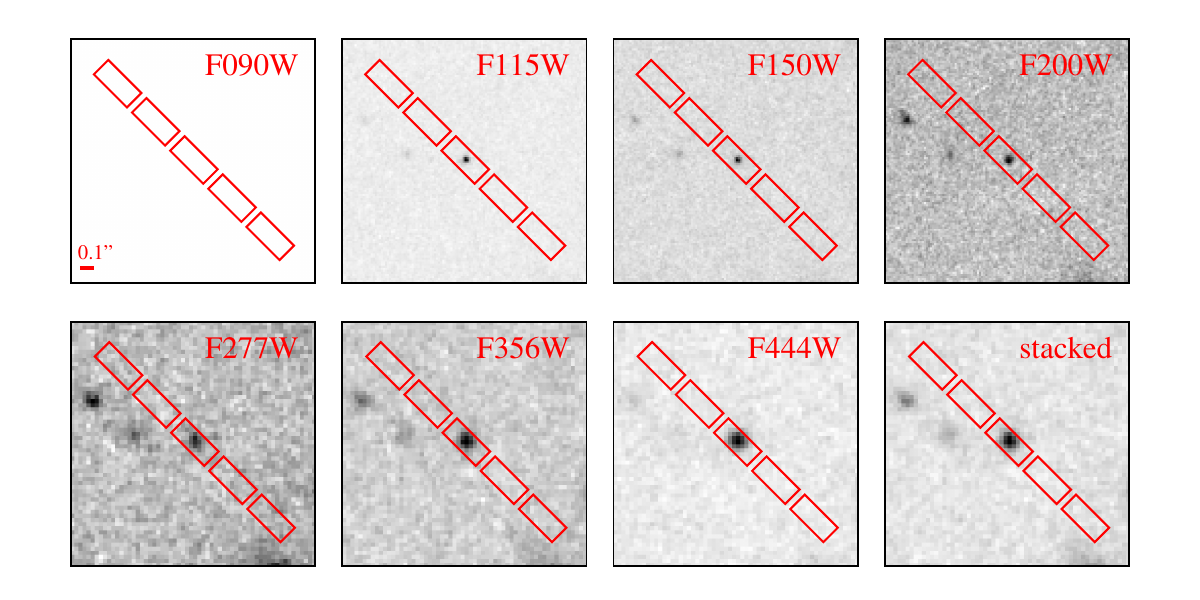}
    \caption{Same as Figure \ref{fig: photometry BD1}, but showing the photometry of Abell2744-BD2.}
    \label{fig: photometry BD2}
\end{figure*}

\begin{figure*}
    \centering
    \includegraphics[width=17cm]{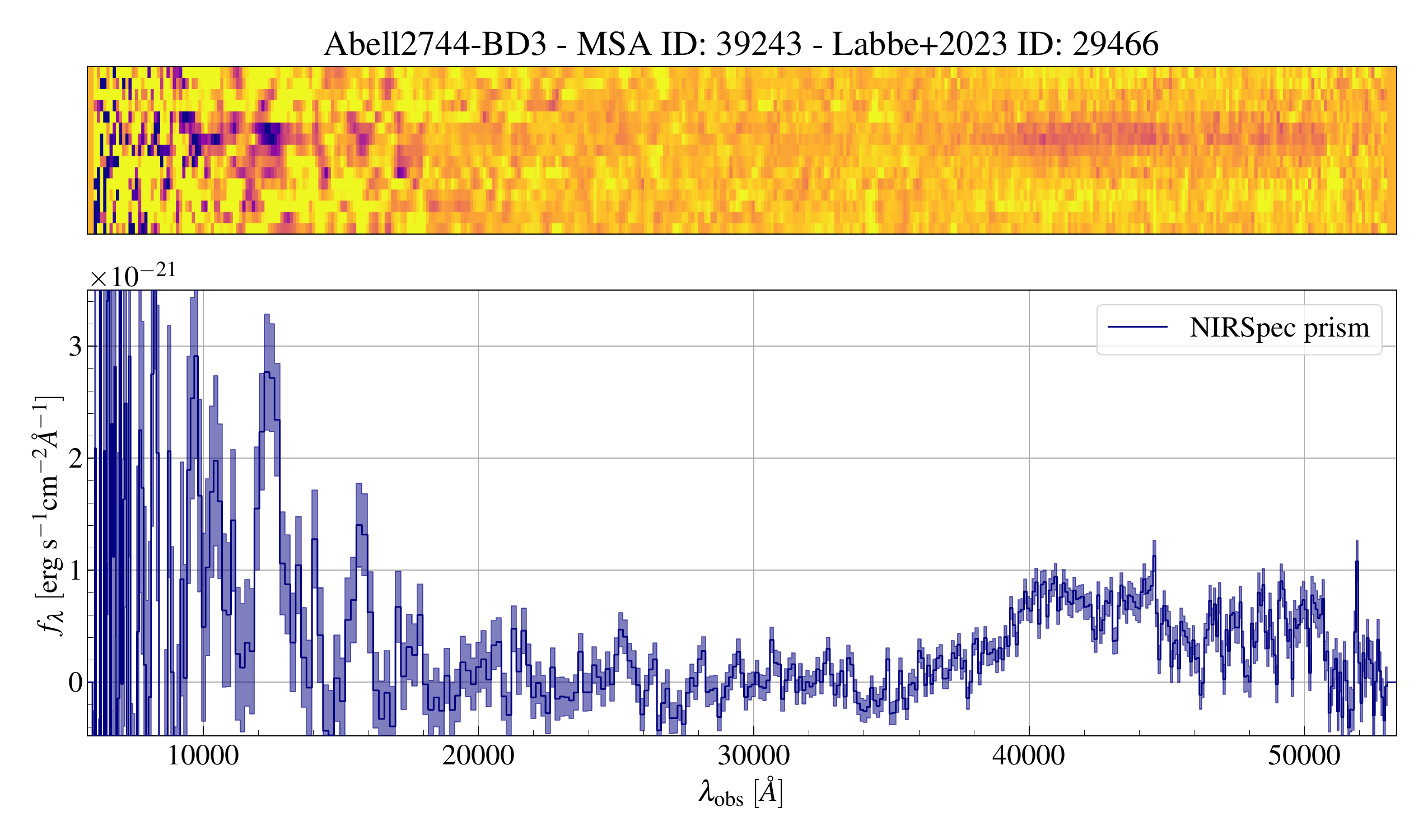}
    \caption{2D spectrum and the extracted 1D spectrum of the brown dwarf Abell2744-BD3.}
    \label{fig: spectrum BD3}
\end{figure*}

\begin{figure*}
    \centering
    \includegraphics[width=17cm]{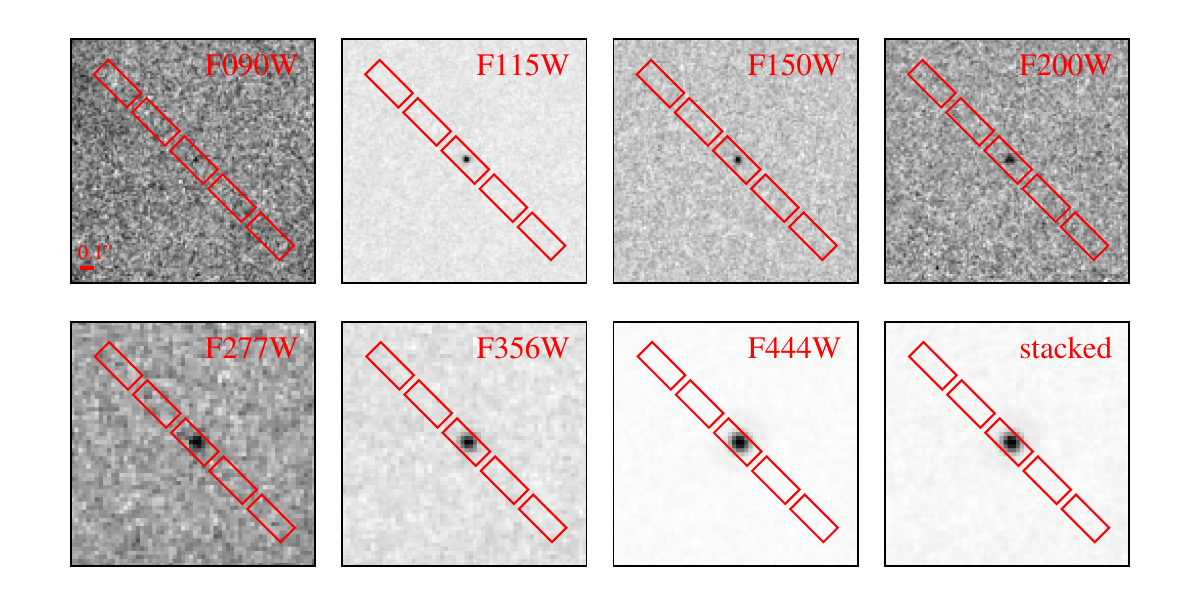}
    \caption{Same as Figures \ref{fig: photometry BD1} and \ref{fig: photometry BD2}, but showing the photometry of Abell2744-BD3.}
    \label{fig: photometry BD3}
\end{figure*}

\begin{figure*}
    \centering
    \includegraphics[width=16cm]{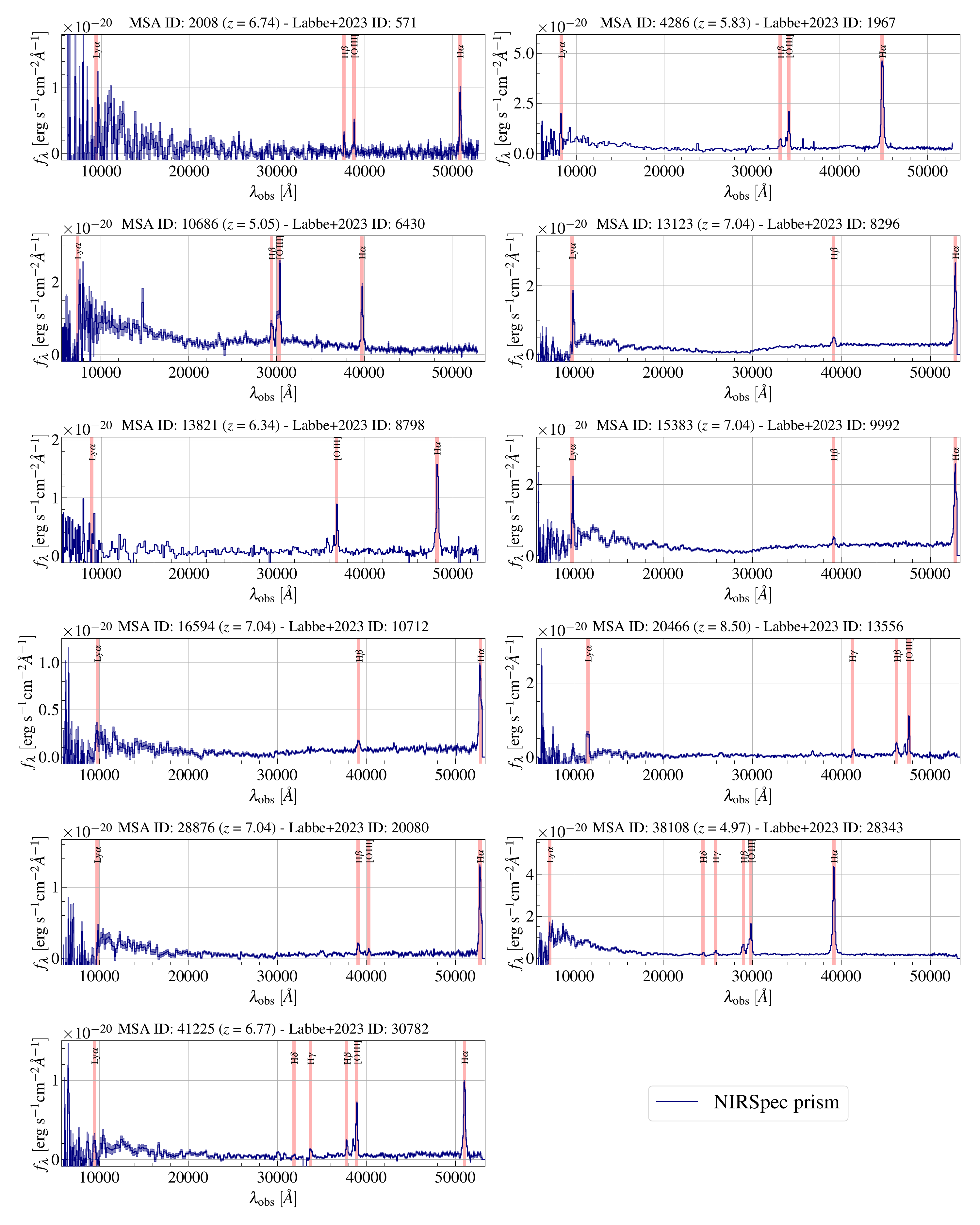}
    \caption{Apart from the three brown dwarfs presented in Figures \ref{fig: spectrum BD1}-\ref{fig: photometry BD3}, we confirm the remaining 11 targeted little red dots as extragalactic sources. For each spectrum, the MSA ID, spectroscopic redshift, and the ID in \cite{Labbe+2023} are noted on top of its corresponding panel. High S/N emission lines are marked with red stripes. MSA IDs 13123, 15383, and 16594 present the triply imaged AGN Abell2744-QSO1 at $z_{\rm spec}=7.05$ \citep{2023ApJ...952..142F, 2023arXiv230805735F}. MSA ID 20466 was recently confirmed as a broad line AGN at $z_{\rm spec}=8.50$ \citep{Kokorev+2023}. The entire sample of these extragalactic sources was more recently analyzed in detail in \cite{greene+2023}, further confirming MSA IDs 4286, 13821, 38108, and 41225 as broad-line AGNs.}
    \label{fig: AGNs}
\end{figure*}

\cite{Labbe+2023} identified a sample of 26 little red dots in a 45 arcmin$^2$ NIRCam coverage of the Abell 2744 field \citep[the UNCOVER program;][]{Bezanson+2022}. These objects were color selected based on their V-shaped spectra, to be blue in the rest-UV (1216--3500\AA) and red at rest-optical ($\lambda > 4000$\AA) wavelengths. They were selected to be dominated by a point-source-like central component. Through spectral energy distribution fitting (SED fitting), a subsample of 17 sources was identified as being dominated by an AGN component at the reddest end of the spectrum (i.e., the F444W photometry). 14 out of these 17 objects were recently followed up by deep NIRSpec multi-object prism spectroscopy through the UNCOVER program. 

We reduced the NIRSpec prism data for the 14 targeted little red dots using the pipeline described in detail in \cite{Langeroodi+2023}. In brief, we converted the raw data into count-rate images (the stage 1 reduction) using the 1.10.0 version of the official STScI JWST pipeline \citep{2022A&A...661A..81F, bushouse_howard_2023_7795697} and the \texttt{jwst\_1106.pmap} Calibration Reference Data System (CRDS) context file. We used the \texttt{msaexp} software \citep{msaexp} for the 2nd and 3rd reduction levels. This includes correcting the residual $1/f$ noise, removing the ``snowball" artifacts, WCS registration, flat-fielding, slit path-loss correction, flux calibration, background subtraction, and extracting the optimal 1D spectra. The extracted spectra of these objects are shown in Figures \ref{fig: spectrum BD1}, \ref{fig: spectrum BD2}, \ref{fig: spectrum BD3}, and \ref{fig: AGNs}. 

Visual inspection confirms three out of 14 objects as brown dwarfs (Abell2744-BD1, Abell2744-BD2, and Abell2744-BD3); the spectra of these sources are shown in Figures \ref{fig: spectrum BD1}, \ref{fig: spectrum BD2}, and \ref{fig: spectrum BD3}. All three objects exhibit the characteristic three-peak features around $1\mu$m, primarily associated with H$_2$O and CH$_4$ absorption. In Section \ref{sec: brown dwarfs} we infer the physical properties of these objects through SED-fitting. We note that Abell2744-BD3 was previously identified as a brown dwarf candidate in NIRCam photometry \citep{2023ApJ...942L..29N}. This object was later identified as a potential highly obscured AGN \citep[see the initial arXiv submission of][]{Labbe+2023}, and hence followed up by UNCOVER's deep NIRSpec MOS.

We confirm the remaining 11 objects (shown in Figure \ref{fig: AGNs}) as extragalactic sources at $z_{\rm spec} \gtrsim 5$ through the identification of Ly$\alpha$ breaks and/or emission lines such as Ly$\alpha$, H$\beta$, [\ion{O}{3}]4959,5007$\lambda\lambda$, or H$\alpha$. It is worth noting that three of these extragalactic sources (MSA IDs 13123, 15383, 16594) are the three images of the triply imaged AGN Abell2744-QSO1 at $z_{\rm spec}=7.05$ \citep{2023ApJ...952..142F, 2023arXiv230805735F}. Moreover, MSA ID 20466 was recently confirmed as a broad line AGN at $z_{\rm spec}=8.50$ \citep{Kokorev+2023}. The nature of the entire sample of these 11 galaxies is discussed in detail in \cite{greene+2023}, where 6 of them are confirmed as broad-line AGNs.

We measured the multi-band HST and NIRCam photometry of the 14 targeted little red dots using the pipeline detailed in \cite{Langeroodi+2023}. The reduced images were acquired from the Grizli Image Release (v6.0) repository\footnote{\url{https://grizli.readthedocs.io/en/latest/grizli/image-release-v6.html}}. Using empirical PSFs, we PSF-matched all the images to the F444W imaging. This excludes the HST WFC3\_IR filters which generally have PSF FWHMs larger than that of F444W; for these filters, we adopted the correction factors measured in \cite{2022ApJ...928...52F} \citep[derived through source injection simulations; see also][]{ceers1}. 

Source photometry was measured in $0.3\arcsec$ circular apertures (diameter) using \texttt{Source Extractor} \citep{sex} in dual image mode. We set \texttt{DETECT\_MINAREA} = 5, \texttt{DETECT\_THRESH} = 3.0, \texttt{DEBLEND\_NTHRESH} = 32, and \texttt{DEBLEND\_MINCOUNT} = 0.005. Aperture correction was performed by scaling all the measured \texttt{MAG\_APER} magnitudes by the difference between the \texttt{MAG\_AUTO} and \texttt{MAG\_APER} flux in the F444W filter \citep[for more details, see][]{Langeroodi+2023}. We corrected for Galactic extinction of the extragalactic sources using the reddening values from \cite{2011ApJ...737..103S} and assuming a \cite{CCM89} attenuation curve with $R_{\rm V} = 3.1$. We corrected the lensing magnification for the extragalactic sources adopting the \cite{2023MNRAS.523.4568F} lens model. Figures \ref{fig: photometry BD1}, \ref{fig: photometry BD2}, and \ref{fig: photometry BD3} show $2.4\arcsec \times 2.4\arcsec$ cutouts of the NIRCam imaging of the three brown dwarfs; their extracted photometry are reported in Table \ref{table: photometry}. The colors of all the targeted little red dots are discussed in Section \ref{sec: contaminants}.

\section{Characterizing the Brown Dwarfs} \label{sec: brown dwarfs}

\begin{figure*}
    \centering
    \includegraphics[width=16cm]{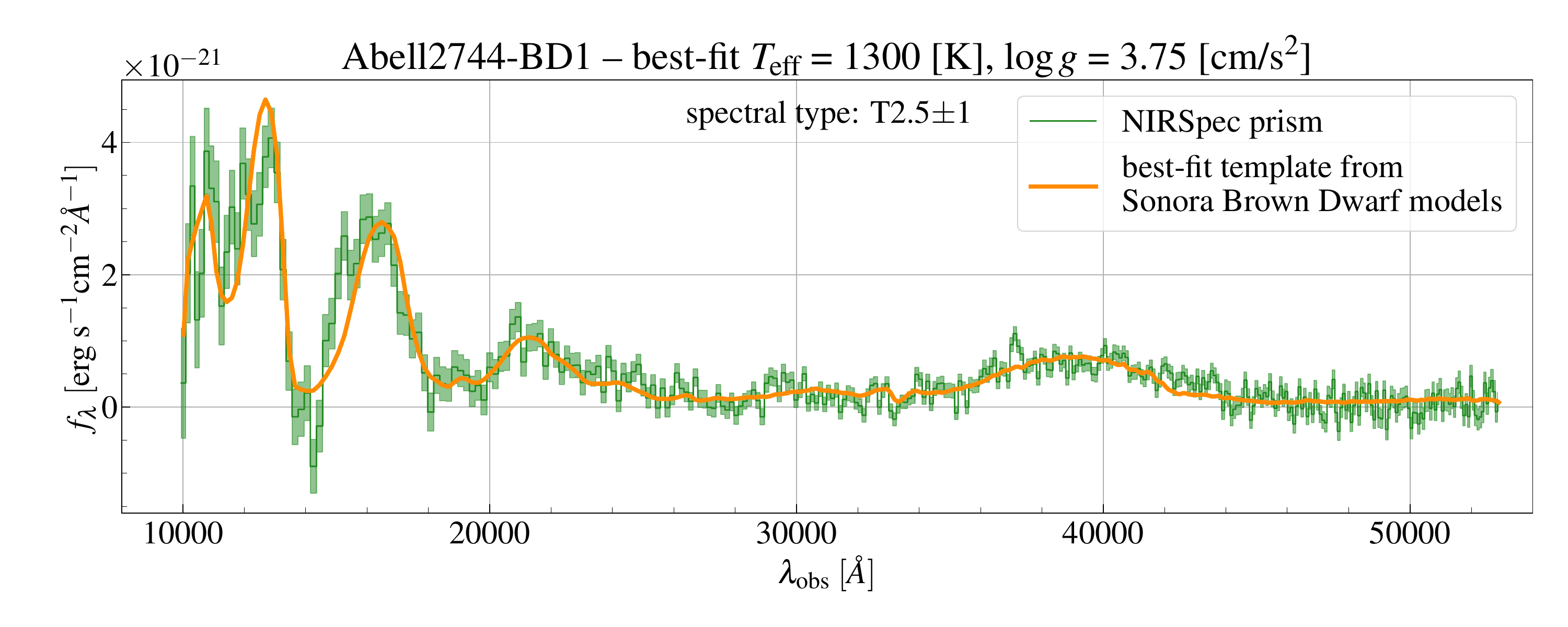}
    \caption{Best-fit model to Abell2744-BD1. The green line shows the observed NIRSpec prism spectra of Abell2744-BD1, and the orange line shows the best-fit template from the grid of Sonora atmosphere models. The best-fit model parameters as well as the corresponding spectral classification are noted at the top of the Figure.}
    \label{fig: BD1}
\end{figure*}

\begin{figure*}
    \centering
    \includegraphics[width=16cm]{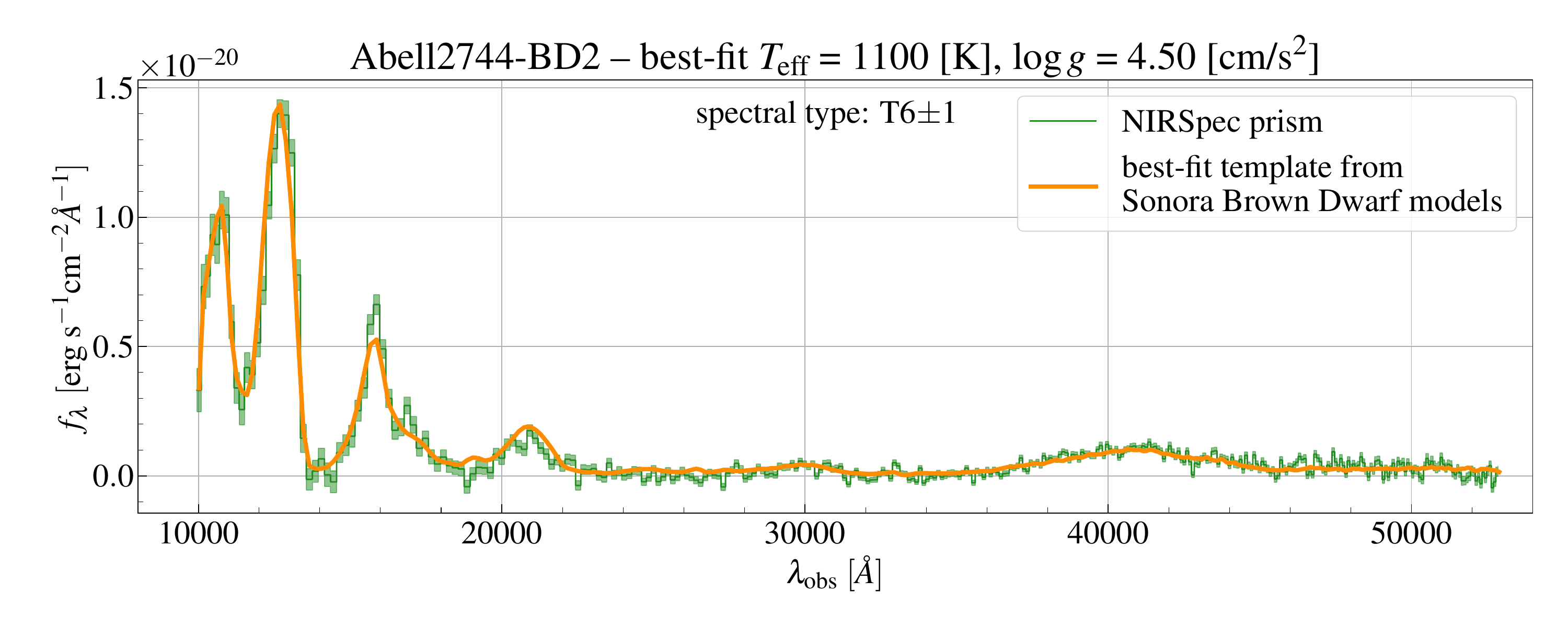}
    \caption{Same as Figure \ref{fig: BD1}; showing the best-fit model to Abell2744-BD2.}
    \label{fig: BD2}
\end{figure*}

\begin{figure*}
    \centering
    \includegraphics[width=16cm]{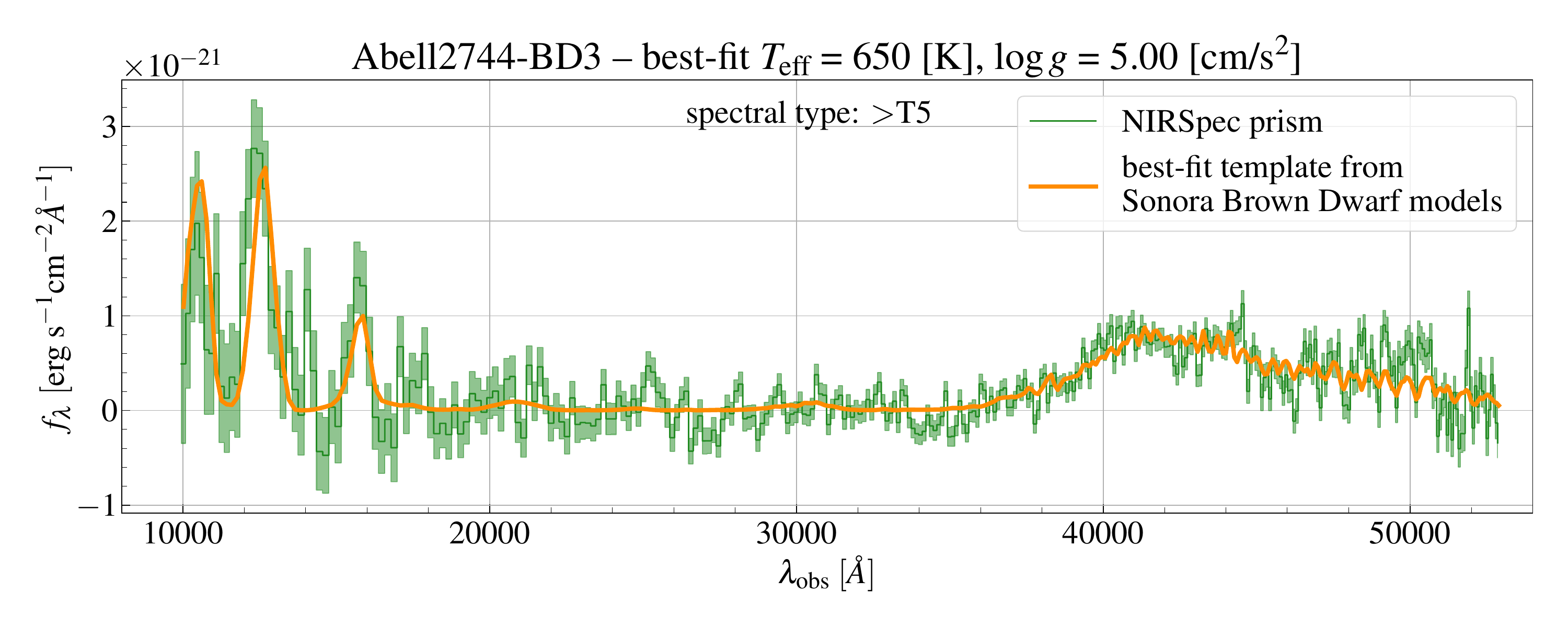}
    \caption{Same as Figures \ref{fig: BD1} and \ref{fig: BD2}; showing the best-fit model to Abell2744-BD3.}
    \label{fig: BD3}
\end{figure*}

We use the Sonora brown dwarf cloud-free atmosphere and evolution models \citep{sonora1, sonora2} to fit the SED of the brown dwarfs discovered in the Abell 2744 field and infer their physical properties and distance. We use the solar metallicity grid of atmosphere models from \cite{sonora2} as spectral templates. This grid is parameterized by an effective temperature in the range $T_{\rm eff}$ = 500--1300 K, a surface gravity in $\log g$ = 3--5.5 $\log({\rm cm/s}^{2})$, and an eddy diffusion coefficient of $\log K_{zz}$ = 2, 4, or 7 $\log ({\rm cm}^{2}/{\rm s})$. This corresponds to 459 unique models. 

\begin{deluxetable}{llll}
\tablewidth{0pt}
\tablecaption{Spectral types and best-fit parameters for the brown dwarfs found in the Abell 2744 field}
\label{table: brown dwarfs}
\tablehead{
\colhead{parameter} &
\colhead{{\scriptsize Abell2744-BD1}} &
\colhead{{\scriptsize Abell2744-BD2}} &
\colhead{{\scriptsize Abell2744-BD3}\tablenotemark{b}}
}
\startdata
MSA ID & 32265 & 33437 & 39243 \\
RA\tablenotemark{c} & 3.537529 & 3.546420 & 3.513891 \\ 
Dec\tablenotemark{c} & $-30.370169$ & $-30.366245$ & $-30.356024$ \\
type & T2.5$\pm$1 & T6$\pm$1 & $>$T5 \\
$T_{\rm eff} [K]$ & $1300\pm50$ \tablenotemark{a} & $1100\pm50$ \tablenotemark{a} & $650\pm50$ \tablenotemark{a} \\
$\log g {\rm [cm/s}^2]$ & $3.75\pm0.25$ \tablenotemark{a} & $4.50\pm0.25$ \tablenotemark{a} & $5.00\pm0.25$ \tablenotemark{a} \\
$\log K_{zz} {\rm [cm}^2/{\rm s}]$ & $7^{+0}_{-3}$ \tablenotemark{a} & $2^{+2}_{-0}$ \tablenotemark{a} & $2^{+2}_{-0}$ \tablenotemark{a} \\
$M/M_{\odot}$ & $0.0047^{+0.0003}_{-0.0002}$ & $0.0157^{+0.0002}_{-0.0003}$ & $0.0305^{+0.0003}_{-0.0001}$ \\
$R/R_{\odot}$ & $0.153^{+0.002}_{-0.002}$ & $0.116^{+0.001}_{-0.001}$ & $0.091^{+0.001}_{-0.000}$ \\
$\log L_{\rm bol}/L_{\odot}$ & $-4.236^{+0.028}_{-0.020}$ & $-4.749^{+0.007}_{-0.009}$ & $-5.875^{+0.004}_{-0.002}$ \\
$\log {\rm Age} [{\rm Gyr}]$ & $6.741^{+0.070}_{-0.055}$ & $8.234^{+0.047}_{-0.050}$ & $9.614^{+0.037}_{-0.052}$ \\
distance [pc] & $4764^{+568}_{-130}$ & $2967^{+603}_{-101}$ & $755^{+105}_{-77}$ \\
\enddata
\tablenotetext{a}{The upper limits on the uncertainties of $T_{\rm eff}$, $\log g$, and $\log K_{zz}$ are dictated by the grid resolution of brown dwarf templates used for the template fitting.}
\tablenotetext{b}{This is the brown dwarf discovered in NIRCam photometry by \cite{2023ApJ...942L..29N}.}
\tablenotetext{c}{J2000.0 (deg)}
\end{deluxetable}

For each brown dwarf discovered in this work, we find the model template that best fits its spectrum. We convolve and resample each model template to match the spectral resolution and sampling of the NIRSpec prism. For each brown dwarf and model template, we use the \texttt{scipy.optimize.curve\_fit} library \citep{scipy} to derive the scaling factor that re-scales the model template to the observed spectrum. For each brown dwarf, we calculate the Gaussian $\chi^2$ of all the re-scaled model templates and choose the one with the smallest $\chi^2$ as the best-fit model. The best-fit models for Abell2744-BD1, Abell2744-BD2, and Abell2744-BD3 are shown as the orange curves in Figures \ref{fig: BD1}, \ref{fig: BD2}, and \ref{fig: BD3}, respectively; their reduced $\chi^2$ values are 1.75, 2.51, and 1.91, respectively.

We use the brown dwarf evolution tables of \cite{sonora1} to convert the $T_{\rm eff}$ and $\log g$ of the best-fit models to physical properties including mass ($M$), radius ($R$), helium mass fraction (Y), bolometric luminosity ($L_{\rm bol}$), and age. For each best-fit $T_{\rm eff}$ and $\log g$, we derive the parameters of interest by considering the full range of metallicities available in the tables. This typically results in 3 distinct values for each parameter of interest, corresponding to the three different metallicities included in the evolution tables. We use these 3 values as the median and $1\sigma$ uncertainty of each parameter of interest. The best-fit parameters are summarized in Table \ref{table: brown dwarfs}. 

\begin{figure}
    \centering
    \includegraphics[width=8.5cm]{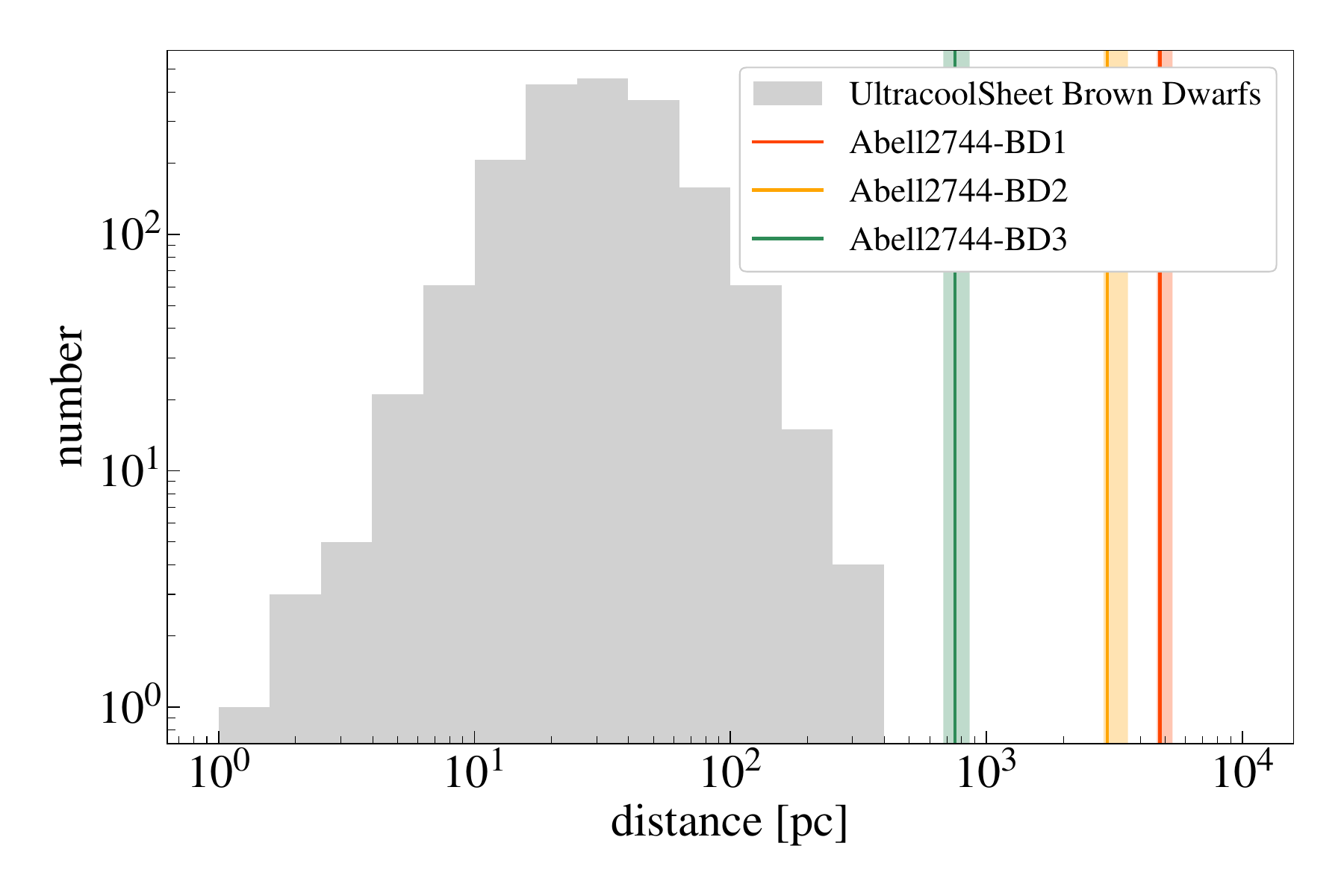}
    \caption{Distances of the brown dwarfs discovered in this work, compared with the UltracoolSheet compilation of literature ultracool brown dwarfs (types late-M and later). The grey histogram presents the distance distribution of a subsample of UltracoolSheet brown dwarfs with reliable parallax distances (see text). The colored vertical lines (shaded regions) indicate the distances ($1\sigma$ uncertainties) of the brown dwarfs discovered in this work. Abell2744-BD1 at $\sim 4.8$ kpc is the most distant brown dwarf discovered to date.}
    \label{fig: distance}
\end{figure}

Following a similar approach, we use the photometry tables of \cite{sonora1} to derive the absolute NIRCam magnitudes for each best-fit model. For each brown dwarf, we calculate one distance per photometry band by comparing the absolute and apparent magnitudes. We report the median and 1$\sigma$ distribution of the distances calculated for each object as its measured distance. Distances are reported in Table \ref{table: brown dwarfs}. 

In Figure \ref{fig: distance} we compare the distances of the brown dwarfs discovered in this work with those from the UltracoolSheet compilation \citep{ultracoolsheet}. UltracoolSheet is a catalog of more than 3000 ultracool brown dwarfs (spectral types of M6 and later) and imaged planets, aiming to provide a complete inventory of all ultracool dwarfs with measured parallaxes. The grey histogram in Figure \ref{fig: distance} shows a subsample of UltracoolSheet with accurate parallaxes (i.e., a parallax S/N of better than $2\sigma$). All three brown dwarfs discovered in this work are at distances farther than what was probed prior to the launch of JWST. Interestingly, Abell2744-BD1 at $4.8^{+0.6}_{-0.1}$ kpc turns out to be the farthest brown dwarf spectroscopically confirmed to date. Such a high distance, at the given high Galactic latitude of this source ($-81.2115$ deg), most likely places this brown dwarf outside the thin disk of the Milky Way \citep[scale height of 0.3 kpc; see also Figure 6 in][and the discussion therein]{hainline+2023}.

We determine the spectral type of each discovered brown dwarf by comparing its spectrum with a spectroscopically uniform library of brown dwarfs with known spectral types. This library is established using the ground-based IRTF/SpeX spectra of $\sim 930$ brown dwarfs, following \citep{2021ApJ...911....7Z}. These template spectra have S/N $> 20$ per pixel in the J-band and are likely single objects (as opposed to unresolved binaries). Each spectral template is scaled to minimize the $\chi^2$ between the science target and the template. The best-matched templates inform the spectral type of the target. 

Figures \ref{fig: BD1 chi2}-\ref{fig: BD3 temp} in Appendix \ref{sec: app1} show the reduced $\chi^2$ as a function of spectral type as well as the best-matching spectral types for the brown dwarfs discovered in this work. Abell2744-BD1: the reduced $\chi^2$-spectral type diagram (Figure \ref{fig: BD1 chi2}) strongly indicates a type T$2.5\pm1$; this is further confirmed through visual inspection of the best-matching templates (Figure \ref{fig: BD1 temp}). Abell2744-BD2: similarly, we classify this sources as a type T$6\pm1$ (Figures \ref{fig: BD2 chi2} and \ref{fig: BD2 temp}). Abell2744-BD3: The spectral type of this brown dwarf appears to be later than the majority of the templates, suggesting a spectral type later than T5 (Figures \ref{fig: BD3 chi2} and \ref{fig: BD1 temp}).

\begin{figure*}
    \centering
    \includegraphics[width=18cm]{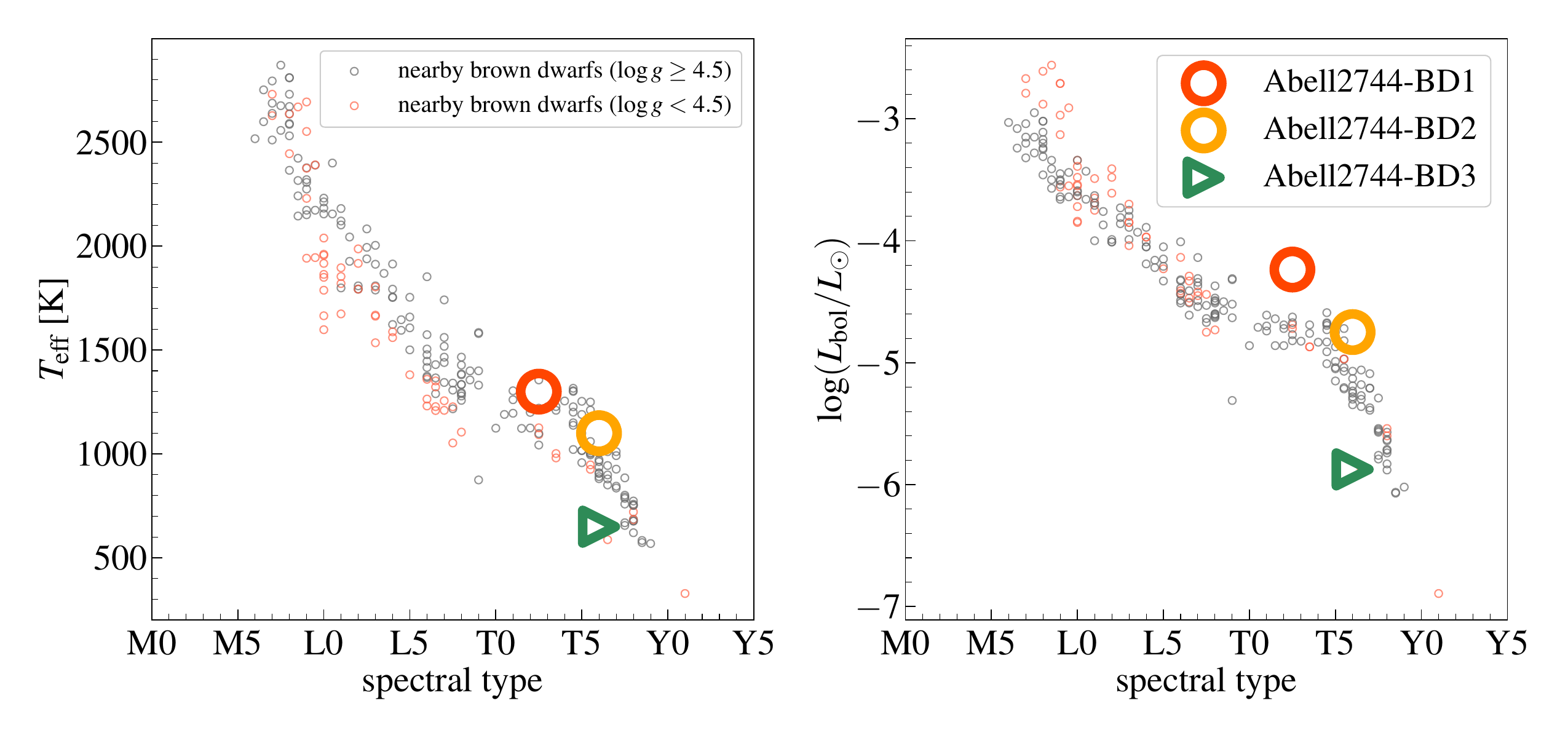}
    \caption{The distant brown dwarfs discovered here have effective temperatures ($T_{\rm eff}$) and bolometric luminosities ($L_{\rm bol}$) consistent with those of benchmark brown dwarfs with similar spectral types. The small data points show the $T_{\rm eff}$-vs-spectral type (left) and $L_{\rm bol}$-vs-spectral type (right) for a compilation of 274 brown dwarfs, including all the 75 known benchmark brown dwarfs of spectral types L6-Y1. The large data points show the locations of Abell2744-BD1 (red), -BD2 (yellow), and -BD3 (green) on these diagrams; the error bars on $T_{\rm eff}$, $L_{\rm bol}$, and spectral type are smaller than the symbol size. We constrain the spectral type of Abell2744-BD3, indicated with the large green data point, to be later than T5.}
    \label{fig: physical properties}
\end{figure*}

We investigate if the physical properties of the distant brown dwarfs discovered in this work are consistent with those of other brown dwarfs of similar spectral types. This is shown in Figure \ref{fig: physical properties}, where we compare the $T_{\rm eff}$ (left) and $L_{\rm bol}$ (right) of Abell2744-BD1, -BD2, and -BD3 with those of a sample of 274 brown dwarfs compiled from \cite{2015ApJ...810..158F}, \cite{2020ApJ...891..171Z}, and \cite{2021ApJ...911....7Z}. Together, \cite{2020ApJ...891..171Z} and \cite{2021ApJ...911....7Z} provide a compilation of all 75 known L6-Y1 benchmark brown dwarfs, sufficiently covering the spectral types of the distant brown dwarfs discovered here. As shown in Figure \ref{fig: physical properties}, our brown dwarfs have effective temperatures and bolometric luminosities consistent with those of the benchmark brown dwarfs with similar spectral types. 

As an independent check we estimated the spectral types based on the relation between effective temperature and spectral type from \cite{2021ApJS..253....7K} (Figure 22 in that study, similar to Figure \ref{fig: physical properties} here), adopting the classification of \cite{2010ApJS..190..100K}. This relation is consistent with our template fitting and suggests that Abell2744-BD1 is an L8-L9 spectral type, close to the border between the L- and T-types; Abell2744-BD2 is a T5 type; and Abell2744-BD3 is a T8-T9 type, close to the border between T- and Y-types.

% We determine the spectral type of each brown dwarf using the trend between effective temperature and spectral type from \cite{2021ApJS..253....7K} (Figure 22 in that study), adopting the classification of \cite{2010ApJS..190..100K}. We classify Abell2744-BD1 with $T_{\rm eff} = 1300$K as an L8-L9 spectral type, close to the border between the L- and T-types; Abell2744-BD2 with $T_{\rm eff} = 1100$K as a T5 type; and Abell2744-BD3 with $T_{\rm eff} = 650$K as a T8-T9 type, close to the border between T- and Y-types. The spectral types are summarized in Table \ref{table: brown dwarfs}. 

\section{Brown Dwarfs as Contaminants in NIRCam-Selected AGNs} \label{sec: contaminants}

\begin{figure}
    \centering
    \includegraphics[width=7.8cm]{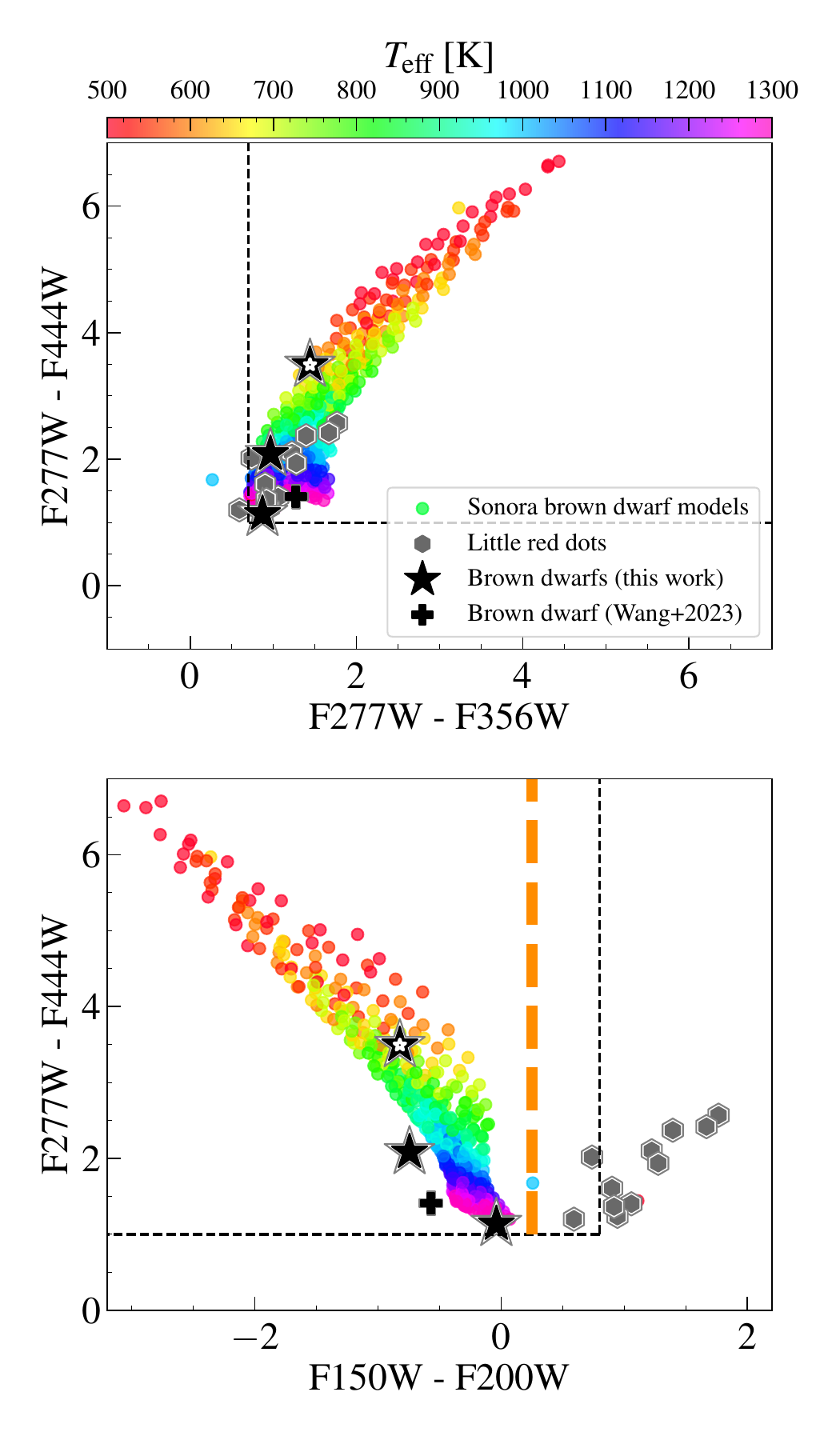}
    \caption{Distribution of brown dwarfs and highly-reddened AGNs on near- and mid-infrared color-color diagrams. Colored circles show the Sonora brown dwarf models, color-coded with their effective temperatures. Grey hexagons show the subsample of highly-reddened AGNs from \cite{Labbe+2023} which we could confirm as extragalactic sources at $z_{\rm spec} \gtrsim 5$. Black stars show the three brown dwarfs presented in this work, masquerading as highly-reddened AGN in the color selection of \cite{Labbe+2023}. The brown dwarf initially discovered in NIRCam photometry by \cite{2023ApJ...942L..29N} is marked with a small white star at its center (see Section \ref{sec: data} for more details). The black plus sign shows the brown dwarf discovered in CEERS NIRCam photometry \citep{2023MNRAS.523.4534W}. The thin dashed black lines show the color selection criteria of \cite{Labbe+2023}. The top panel shows that as brown dwarfs become warmer, they resemble the mid-infrared colors of highly-reddened AGNs more closely. The bottom panel shows that the brown dwarfs are generally much bluer than highly-reddened AGNs in the near-infrared filters. This is evident by the horizontal separation between the brown dwarfs and highly-reddened AGNs in this panel. We propose a cut at F150W$-$F200W = 0.25 mag to distinguish brown dwarfs from highly-reddened AGNs; this is shown as the thick dashed orange line.}
    \label{fig: cc}
\end{figure}

As shown in Section \ref{sec: data}, three out of the 14 targeted NIRCam-selected little red dots turned out to be brown dwarfs. Moreover, another three spectra correspond to the triply imaged AGN Abell2744-QSO1  \citep{2023ApJ...952..142F, 2023arXiv230805735F}. This results in a 3/12 $=$ 25\% brown dwarf contamination fraction in NIRCam-selected highly reddened AGN, if the color selection criteria of \cite{Labbe+2023} are adopted \footnote{Either ``criteria 1'' = (F115W--F150W $<0.8$mag) \& (F200W--F277W $>0.7$mag) \& (F200W--F356W $>1.0$mag); or ``criteria 2'' = (F150W--F200W $<0.8$mag) \& (F277W--F356W $>0.7$mag) \& (F277W--F444W $>1.0$mag). These selection criteria are shown in Figure \ref{fig: cc} as the thin dashed lines.}.

The reason for this high contamination rate becomes clear when considering the distribution of brown dwarfs and highly-reddened AGN in the F277W$-$F444W vs.\ F277W$-$F356W color-color diagram. This is shown in the top panel of Figure \ref{fig: cc}, where the Sonora brown dwarf models are color-coded with their effective temperatures and the NIRCam-selected highly-reddened AGN candidates from \cite{Labbe+2023} (that did not turn out to be brown dwarfs) are shown as grey hexagons (we are only showing the sources which were spectroscopically followed-up). The dashed lines indicate the selection criteria of \cite{Labbe+2023}. As brown dwarfs become warmer, they resemble more closely the location of little red dots in this diagram. 

The top panel of Figure \ref{fig: cc} shows that it would be difficult to distinguish between the brown dwarfs warmer than $\sim 800$ K and highly-reddened AGN solely based on the F277W$-$F444W vs.\ F277W$-$F356W diagram. This is the case for two of the brown dwarfs discovered in this work with effective temperatures above 1100 K, as shown by the black stars. However, since the $4 \mu$m feature in the spectra of brown dwarfs becomes stronger with decreasing temperature, the extremely red mid-infrared colors of brown dwarfs at effective temperatures below 700 K become difficult to replicate by highly-reddened AGN. This can help minimize the contamination of $T_{\rm eff} < 700$ brown dwarfs in the NIRCam selection of little red dots. 

The bottom panel in Figure \ref{fig: cc} shows the distribution of brown dwarfs and little red dots on the F150W$-$F200W vs.\ F277W$-$F444W color-color diagram. Like the top panel, as brown dwarfs become warmer they resemble the colors of the little red dots more closely. However, brown dwarfs never seem to match the colors of the little red dots in this diagram: in the near-infrared filters, brown dwarfs are in general much bluer than the highly-reddened AGN. This can be leveraged to minimize the brown dwarf contamination in color-selected samples of highly obscured AGNs. The AGN selection criteria \citep[e.g., like those of][]{Labbe+2023} can be supplemented with a simple color cut at F150W$-$F200W = 0.25 mag to distinguish the brown dwarfs from highly-reddened AGN; objects with F150W$-$F200W $<$ 0.25 mag should be classified as brown dwarfs. This color cut is shown as the dashed orange line in Figure \ref{fig: cc}.

The above selection criteria seem more intuitive in the context of the inferred UV slopes of color-selected highly-reddened AGNs. If brown dwarfs are misclassified as high-redshift sources, their extremely blue near-infrared colors (F150W$-$F200W $<$ 0.25 mag) can yield UV slopes that are much steeper than $-3$. For instance, Abell2744-BD2 was classified as a $z \sim 7.5$ source with a UV slope of $-4$ \citep{Labbe+2023}. Such steep slopes are widely considered unphysical, as they cannot be reproduced even with un-reddened spectra (i.e., 100\% escape fraction) of extremely young and metal-poor stars \citep[see, e.g.,][]{2022ApJ...941..153T}.

\section{Conclusion} \label{sec: conclusion}

We analyzed the NIRSpec multi-object prism follow-ups of a NIRCam-selected sample of 14 candidate highly-reddened AGNs. We identified three sources as brown dwarfs with effective temperatures between 650 and 1300 K and distances between 0.8 and 4.8 kpc. Abell2744-BD1 at 4.8 kpc is the most distant brown dwarf discovered to date; its distance and Galactic latitude place it out of the thin disk of the Milky Way. We confirm the remaining 11 objects as extragalactic sources at $z_{\rm spec} \gtrsim 5$. This corresponds to a 25\% brown dwarf contamination fraction in NIRCam-selected samples of little red dots. The NIRCam photometry shows that brown dwarfs are in general much bluer than highly-reddened AGNs in near-infrared filters. This indicates that a simple near-infrared color cut at F150W$-$F200W = 0.25 mag can minimize the brown dwarf contamination in NIRCam selection of highly-reddened AGNs: objects with  F150W$-$F200W $<$ 0.25 mag should be classified as brown dwarfs.

Following the submission of this work, an independent analysis conducted by \cite{2023arXiv230812107B} appeared, in general agreement with our findings. We also note that following the submission of this work, \cite{2023arXiv230905835H} reported the spectroscopic discovery of a candidate brown dwarf in CEERS NIRSpec prism spectroscopy (MSA ID 1558). We reduced the CEERS data \citep[see][]{Langeroodi+2023} and find that the candidate appears to be substantially hotter than the maximum temperature of the Sonora brown dwarf models (1300 K).

\software{
msaexp \citep{msaexp},
Source Extractor \citep{sex},
sedpy \citep{sedpy},
scipy \citep{scipy}
}

\section*{Acknowledgments}

We express our gratitude to the UNCOVER team (PIs: Labbe and Bezanson) for designing the NIRCam survey of the Abell 2744 field, the initial selection of the red compact sources, and conducting the NIRSpec follow-ups. We are indebted to Zhoujian Zhang for his swift assistance with spectral typing the brown dwarfs and his guidance in preparing Figures~\ref{fig: distance} and \ref{fig: physical properties}. The authors were supported by a VILLUM FONDEN Investigator grant (project number 16599). The JWST data presented in this paper were obtained from the Mikulski Archive for Space Telescopes (MAST) at the Space Telescope Science Institute. The specific observations analyzed can be accessed via \dataset[DOI]{http://dx.doi.org/10.17909/8enx-3310}.

\newpage

\bibliography{main}
\bibliographystyle{aasjournal}

\appendix

\section{Determining the Spectral Types of Brown Dwarfs} \label{sec: app1}

Here we present the reduced $\chi^2$ vs spectral type as well as the best-matching spectral types for the brown dwarfs discovered in this work (see Section \ref{sec: brown dwarfs} for more details).

\begin{figure*}[hb]
    \centering
    \includegraphics[width=12cm]{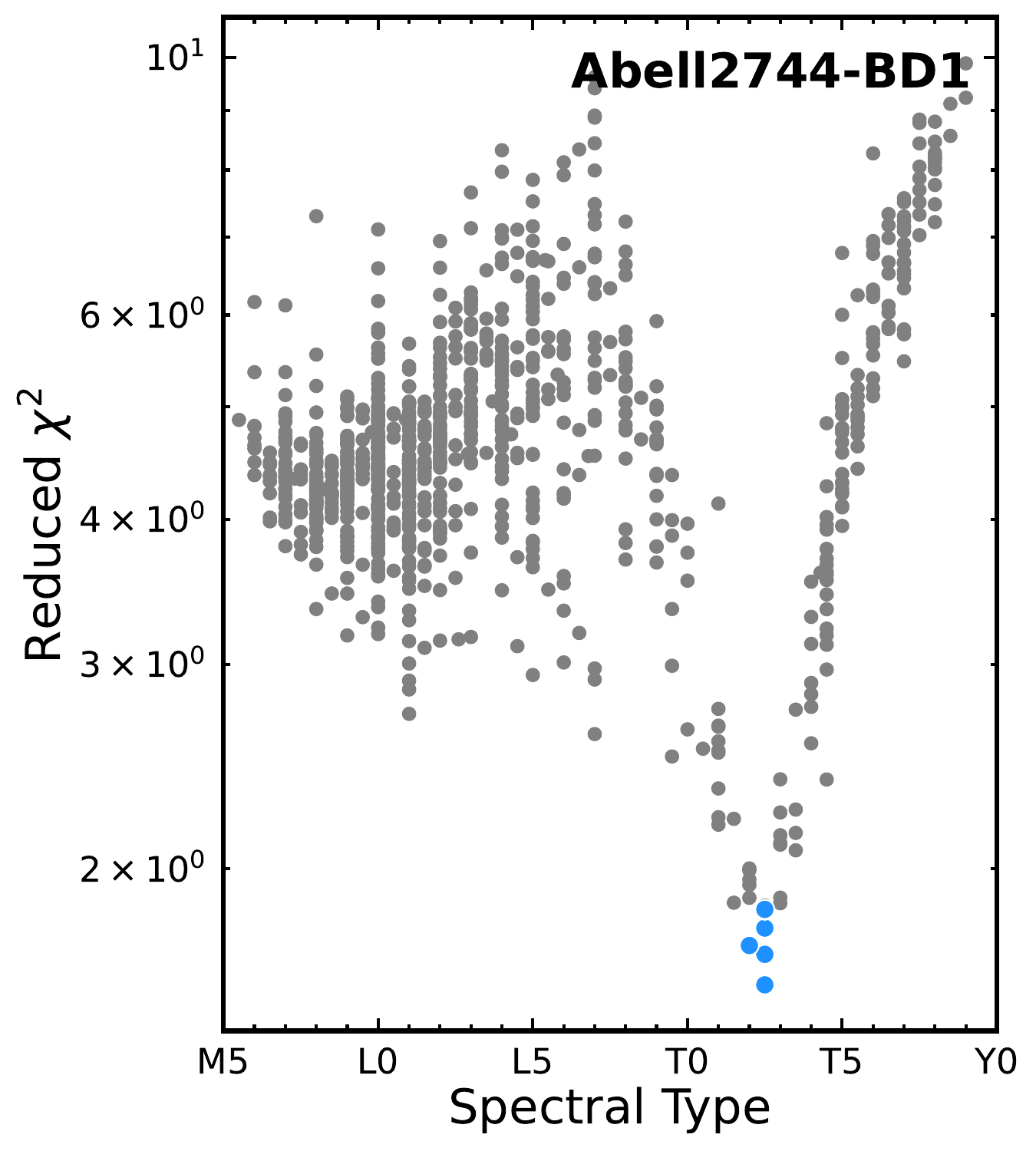}
    \caption{Reduced $\chi^2$ vs spectral type for Abell2744-BD1. The grey data points show the reduced $\chi^2$ for each spectral template. The blue data points show the best-matching templates with the lowest reduced $\chi^2$.}
    \label{fig: BD1 chi2}
\end{figure*}

\begin{figure*}
    \centering
    \includegraphics[width=7.5cm]{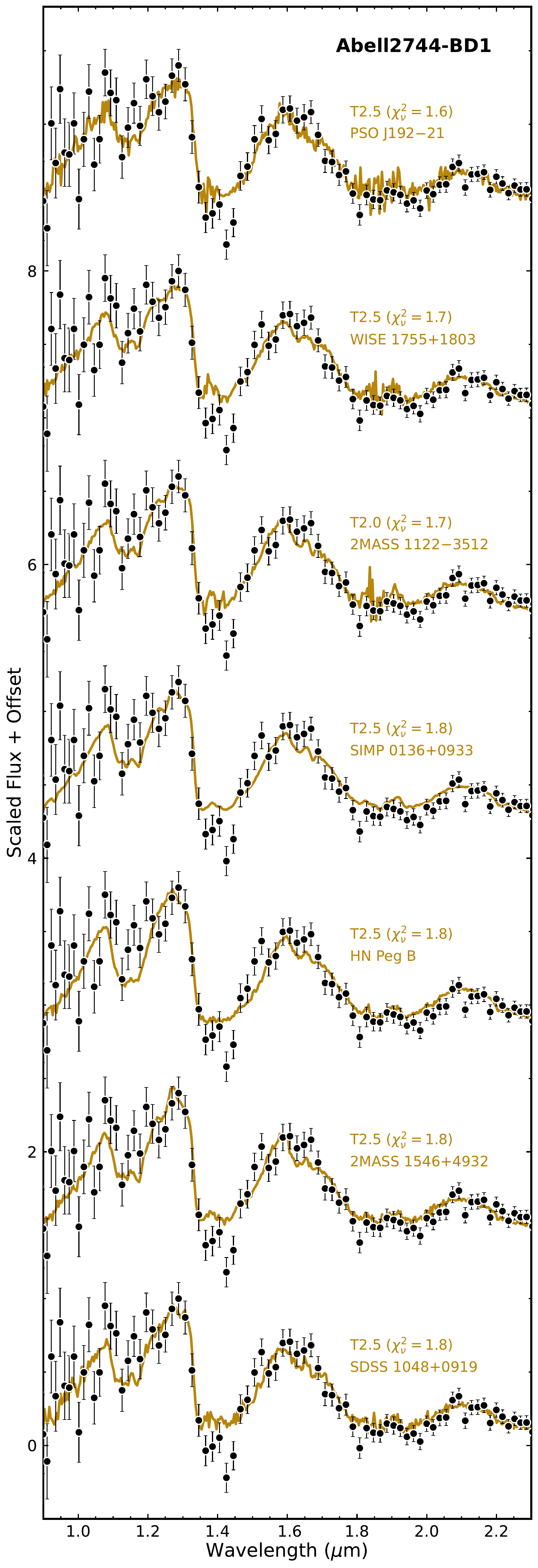}
    \caption{Best-matching templates (yellow lines) to the brown dwarf Abell2744-BD1 (black data points). The $\chi^2$ of these templates are shown as the blue data points in Figure \ref{fig: BD1 chi2}.}
    \label{fig: BD1 temp}
\end{figure*}

\begin{figure*}
    \centering
    \includegraphics[width=12cm]{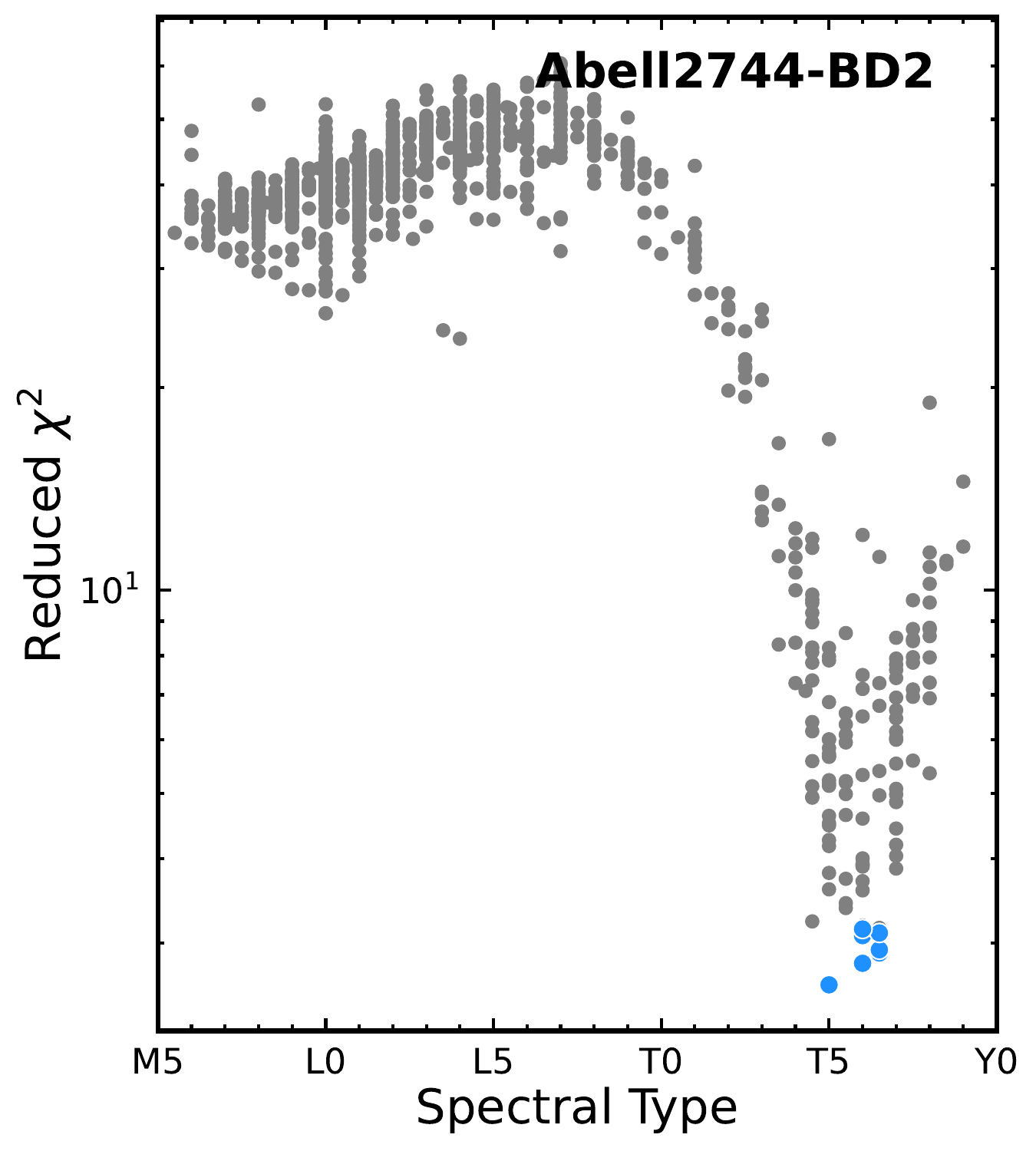}
    \caption{Same as Figure \ref{fig: BD1 chi2}, showing the reduced $\chi^2$ vs spectral type for Abell2744-BD2.}
    \label{fig: BD2 chi2}
\end{figure*}

\begin{figure*}
    \centering
    \includegraphics[width=7.5cm]{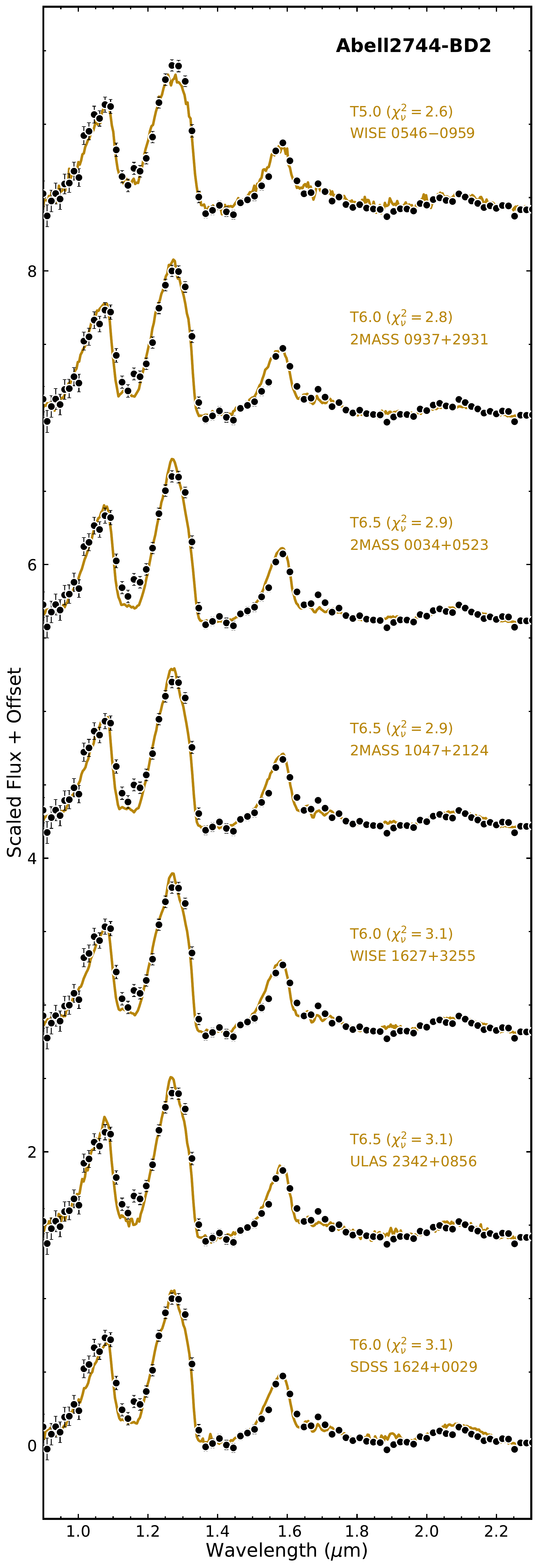}
    \caption{Best-matching templates (yellow lines) to the brown dwarf Abell2744-BD2 (black data points). The $\chi^2$ of these templates are shown as the blue data points in Figure \ref{fig: BD2 chi2}.}
    \label{fig: BD2 temp}
\end{figure*}

\begin{figure*}
    \centering
    \includegraphics[width=12cm]{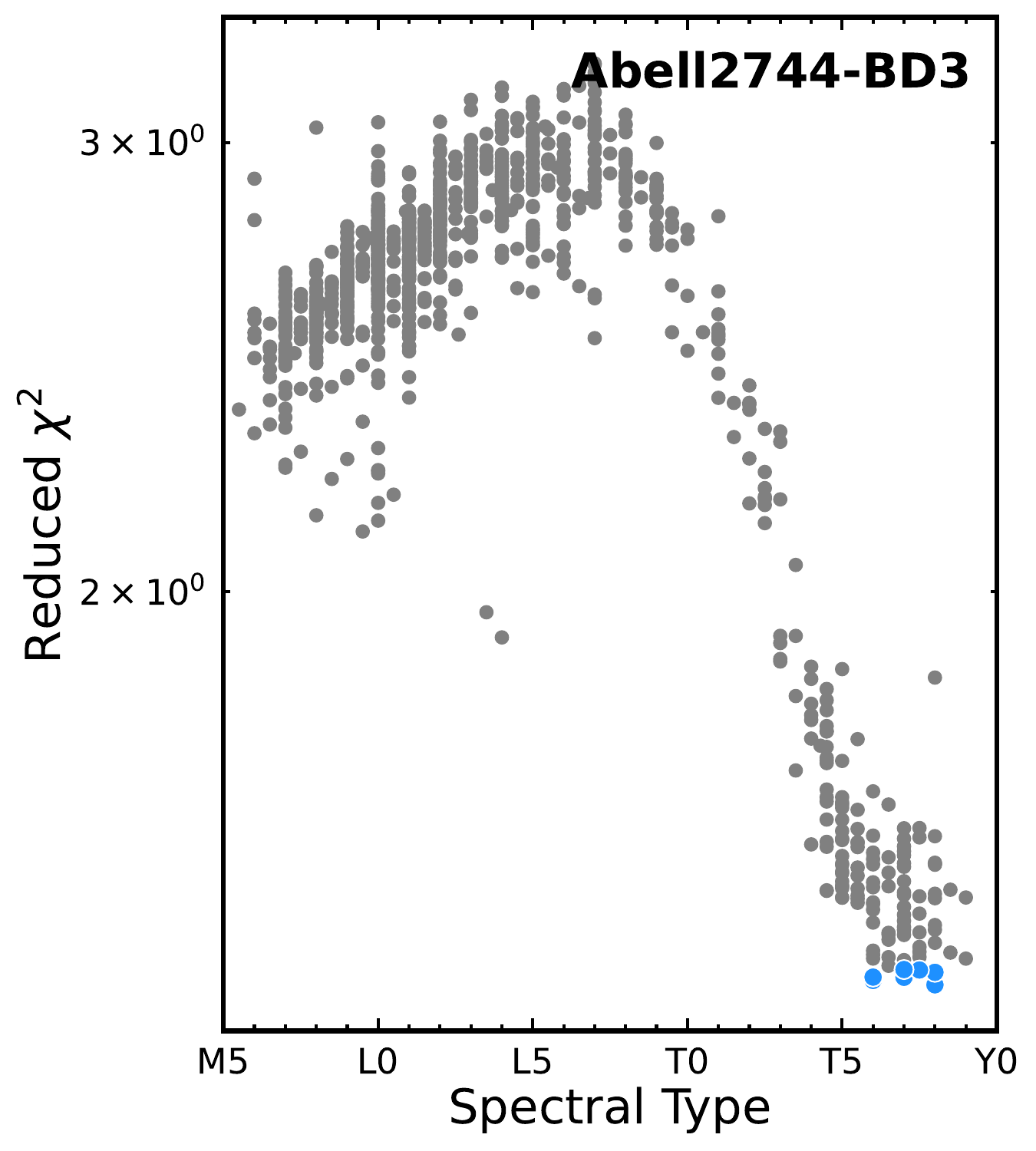}
    \caption{Same as Figure \ref{fig: BD1 chi2}, showing the reduced $\chi^2$ vs spectral type for Abell2744-BD3.}
    \label{fig: BD3 chi2}
\end{figure*}

\begin{figure*}
    \centering
    \includegraphics[width=7.5cm]{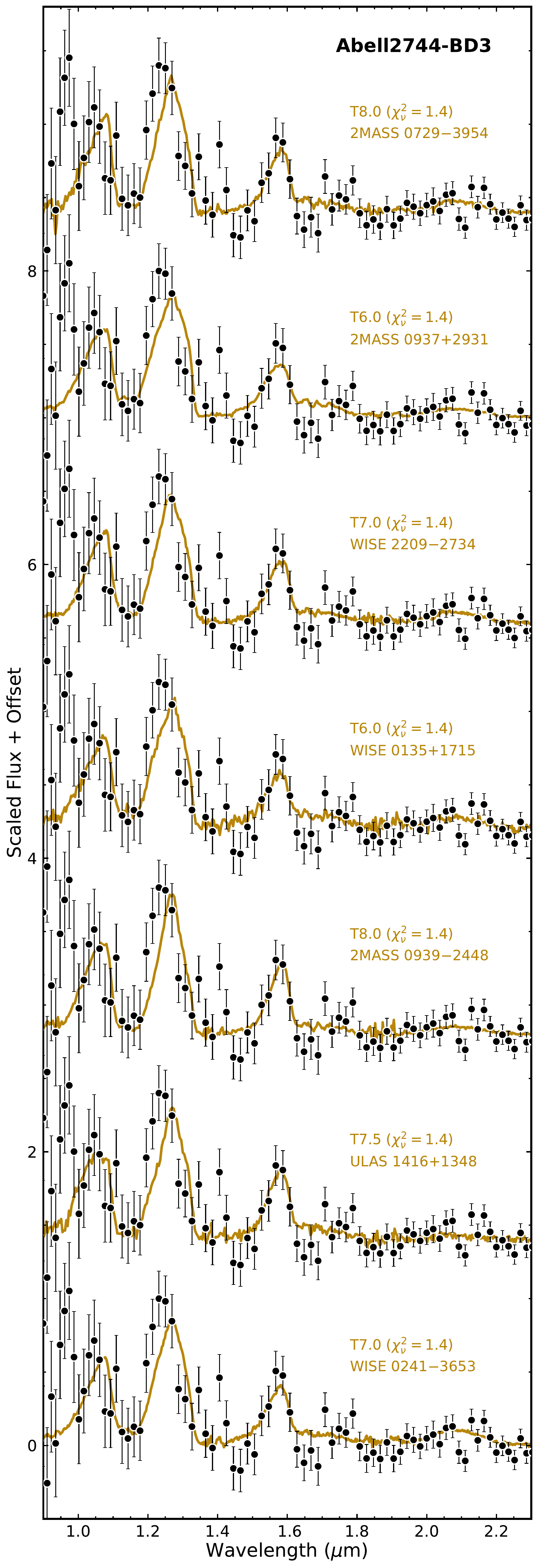}
    \caption{Best-matching templates (yellow lines) to the brown dwarf Abell2744-BD3 (black data points). The $\chi^2$ of these templates are shown as the blue data points in Figure \ref{fig: BD3 chi2}.}
    \label{fig: BD3 temp}
\end{figure*}

\end{document}